\documentclass[runningheads]{llncs}

\usepackage[utf8]{inputenc}

\usepackage{latexsym}

\usepackage{graphicx}
\usepackage{pstricks}
\usepackage{amssymb}
\usepackage{amsmath}

\usepackage{plain}

\input{kempe09a.def}

\begin{document}

\mainmatter

\title{Selected Operations and Applications \\
  of $n$-Tape Weighted Finite-State Machines
  \thanks{
    Sections~\ref{sec:prelim}--\ref{sec:alg} are based on published results
    \cite{kempe+champarnaud+eisner:2004a,kempe+al:2005a,kempe+al:2005b,champarnaud+al:2008a},
    obtained at Xerox Research Centre Europe (XRCE), Meylan, France,
    through joint work between
    Jean-Marc Champarnaud (Rouen Univ.),
    Jason Eisner (Johns Hopkins Univ.),
    Franck Guingne and Florent Nicart (XRCE and Rouen Univ.),
    and the author.
  }%
}

\titlerunning{$n$-Tape Weighted Finite-State Machines}

\author{André Kempe}

\authorrunning{André Kempe}

\institute{
  Cadège Technologies  \\
  18 rue de Vouillé -- 75015 Paris -- France  \\
  \email{http://a.kempe.free.fr} -- \email{a.kempe@free.fr}
}

\maketitle

\begin{abstract}
  A weighted finite-state machine with $n$ tapes ($n$-WFSM)
  defines a rational relation on $n$ strings.
  The paper recalls
  important operations on these relations,
  and an algorithm for their auto-intersection.
  Through a series of practical applications,
  it investigates the augmented descriptive power of $n$-WFSMs,
  w.r.t. classical 1- and  2-WFSMs (acceptors and transducers).
  Some applications are not feasible with the latter.
\end{abstract}


\section{Introduction
  \label{sec:intro}}

A weighted finite-state machine with $n$ tapes ($n$-WFSM)
\cite{rabin+scott:1959,elgot+mezei:1965,kay:1987,harju+karhumaki:1991,kaplan+kay:1994}
defines a rational relation on $n$ strings.
It is a generalization of weighted acceptors (one tape) and transducers (two tapes).

This paper investigates the potential of $n$-ary rational relations (resp. $n$-WFSMs)
compared to languages and binary relations (resp. acceptors and transducers),
in practical tasks.
All described operations and applications have been implemented with
Xerox's WFSC tool~\cite{kempe+al:2003}.

The paper is organized as follows:
Section~\ref{sec:prelim}
recalls some basic definitions about
$n$-ary weighted rational relations and $n$-WFSMs.
Section~\ref{sec:operations}
summarizes some central operations on these relations and machines,
such as join and auto-intersection.
Unfortunately, due to Post's Correspondence Problem,
there cannot exist a fully general auto-intersection algorithm.
Section~\ref{sec:alg}
recalls a restricted algorithm for a class of $n$-WFSMs.
Section~\ref{sec:applic}
demonstrates the augmented descriptive power of $n$-WFSMs
through a series of practical applications, namely
the morphological analysis of Semitic languages (\ref{ssec:semitic}),
the preservation of intermediate results in transducer cascades (\ref{ssec:cascades}),
the induction of morphological rules from corpora (\ref{ssec:morph}),
the alignment of lexicon entries (\ref{ssec:lexalign}),
the automatic extraction of acronyms and their meaning from corpora (\ref{ssec:acronyms}), and
the search for cognates in a bilingual lexicon (\ref{ssec:cognates}).


\section{Definitions
  \label{sec:prelim}}

We recall some definitions about
$n$-ary weighted relations and their machines,
following the usual definitions for multi-tape
automata \cite{elgot+mezei:1965,eilenberg:1974},
with semiring weights added just as for acceptors and transducers
\cite{kuich+salomaa:1986,mohri+al:1998}.
For more details see \cite{kempe+champarnaud+eisner:2004a}.

\smallskip

A {\it weighted $n$-ary relation}\/ is a function from $(\Sigma^*)^n$ to
$\srSetK$, for a given finite alphabet $\Sigma$ and a given weight
semiring $\srK = \aTuple{\srSetK, \srPlus, \srTimes, \srZero,
  \srOne}$.  A relation assigns a weight to any
$n$-tuple of strings.  A weight of $\srZero$ can be interpreted as
meaning that the tuple is not in the relation.
We are especially interested in {\it rational} (or {\it regular})
$n$-ary relations, i.e. relations that can be encoded by $n$-tape
weighted finite-state machines, that we now define.

We adopt the convention that variable names referring to $n$-tuples of
strings include a superscript $\tapnum{n}$.  
Thus we write $s\tapnum{n}$ rather than
${\mathop{s}\limits^\rightarrow}$ 
for a tuple of strings $\aTuple{s_1, \dots  s_n}$.  
We also use this convention for the names of 
objects that contain $n$-tuples of strings,
such as $n$-tape machines and their transitions and paths.

\smallskip

An {\it $n$-tape weighted finite-state machine}\/ ($n$-WFSM)
$A\tapnum{n}$ is defined by a six-tuple
$A\tapnum{n} = \aTuple{\Sigma, Q, \srK, E\tapnum{n}, \wgtInit, \wgtFin}$,
with
$\Sigma$ being a finite alphabet,
$Q$ a finite set of states,
$\srK\!=\!\aTuple{\srSetK,\srPlus,\srTimes,\srZero,\srOne}$ the
semiring of weights, 
$E\tapnum{n}\!\subseteq ( Q\times (\Sigma^*)^n \times \srSetK \times Q )$
		a finite set of weighted $n$-tape transitions, 
$\wgtInit : Q \rightarrow \srSetK$ a function that assigns initial
weights to states, 
and $\wgtFin : Q \rightarrow \srSetK$ a function that assigns final
weights to states.

Any transition $e\tapnum{n}\!\in\!E\tapnum{n}$ has the form
$e\tapnum{n}\!=\!\aTuple{\eSrc,\lab\tapnum{n},w,\eTrg}$.
We refer to these four components as the transition's source state
$\eSrc(e\tapnum{n})\!\in\!Q$, its label
$\lab(e\tapnum{n})\!\in\!(\Sigma^*)^n$, its weight
$w(e\tapnum{n})\!\in\!\srSetK$, and its target state
$\eTrg(e\tapnum{n})\!\in\!Q$.  
We refer by $E(q)$ to the set of out-going transitions of a state $q\!\in\!Q$
~(with $E(q)\!\subseteq\!E\tapnum{n}$).

A {\it path}\/ $\path\tapnum{n}$ of length $k \geq 0$
is a sequence of transitions
$e_1\tapnum{n} e_2\tapnum{n} \cdots e_k\tapnum{n}$
such that $\eTrg(e_i\tapnum{n})\!=\!\eSrc(e_{i+1}\tapnum{n})$
for all $i\!\in\!\aRange{1, k\!-\!1}$.
The label of a path is the element-wise concatenation of 
the labels of its transitions.
The weight of a path $\path\tapnum{n}$ is
\begin{equation}
  w(\path\tapnum{n})	\DefAs
	\wgtInit(\eSrc(e_1\tapnum{n})) \srTimes
	\left(\srBigTimes_{j\in\aRange{1,k}}
		\spc{-1ex}w\left(e_j\tapnum{n}\right)\right) \srTimes
	\wgtFin(\eTrg(e_k\tapnum{n}))
\end{equation}

\noindent
The path is said to be {\it successful}, and to {\it accept}
its label, if $w(\path\tapnum{n})\neq\srZero$.


\section{Operations
	\label{sec:operations}}

We now recall some central operations on $n$-ary weighted relations and $n$-WFSMs
\cite{kempe+al:2004a}.
The auto-intersection operation was introduced,
with the aim of simplifying the computation of the join operation.  
The notation is inspired by relational databases.  
For mathematical details of simple operations see~\cite{kempe+champarnaud+eisner:2004a}.

\subsection{Simple Operations}

Any $n$-ary weighted rational relation
can be constructed by combining
the basic rational operations of {\it union}\/, {\it concatenation}\/ and {\it closure}\/.
Rational operations can be implemented by simple constructions on the
corresponding non-deterministic $n$-tape WFSMs \cite{rosenberg:1964}.
These $n$-tape constructions and their semiring-weighted versions are
exactly the same as for acceptors and transducers, since
they are indifferent to the $n$-tuple transition labels.

The {\em projection} operator $\Proj{j_1,\dots j_m}$,
with $j_1,\dots j_m\!\in\!\aRange{1,n}$,
maps an $n$-ary relation to an $m$-ary one
by retaining in each tuple components
specified by the indices $j_1, \dots j_m$ and
placing them in the specified order.
Indices may occur in any order, possibly with repeats.
Thus the tapes can be permuted or duplicated: 
$\Proj{2,1}$ inverts a 2-ary relation.
The {\it complementary projection}\/ operator $\CProj{j_1,\ldots j_m}$
removes the tapes $j_1,\ldots j_m$ and preserves the order of other tapes.

\subsection{Join operation}\label{sec:ops:mtjoin}

The $n$-WFSM {\it join} operator differs from database join in that
database columns are named, whereas our tapes are numbered.
Since tapes must explicitly be selected by number,
join is neither associative nor commutative.

For any distinct $i_1,\dots i_r\!\in\! \aRange{1,n}$
and any distinct $j_1,\dots j_r\!\in\! \aRange{1,m}$,
we define a {\em join} operator $\JOIN{i_1=j_1,\dots i_r=j_r}$.  It
combines an $n$-ary and an $m$-ary relation into an $(n+m-r)$-ary
relation defined as follows:\footnote{
For example the tuples $\aTuple{abc,def,\epsilon}$ and
$\aTuple{def,ghi,\epsilon,jkl}$ combine in the join 
$\JOIN{2=1,3=3}$
and yield the tuple $\aTuple{abc,def,\epsilon,ghi,jkl}$, with
a weight equal to the product of their weights.}
\begin{equation}
  \spc{-2.5ex}
  \scalebox{0.85}[1.0]{$
    \left(\relat_1\tapnum{n} \JOIN{i_1=j_1,\dots i_r=j_r}
    \relat_2\tapnum{m}\right)(\aTuple{u_1,\dots u_n,s_1,\dots s_{m-r}})
    \DefAs \relat_1\tapnum{n}(u\tapnum{n}) \srTimes
    \relat_2\tapnum{m}(v\tapnum{m})
  $} %
\end{equation}

\noindent 
$v\tapnum{m}$ being the unique tuple s. t.
\scalebox{0.9}[1.0]{$\CProj{j_1,\ldots j_r}(v\tapnum{m})=s\tapnum{m-r}$} and 
\scalebox{0.9}[1.0]{$(\forall k \in \aRange{1,r})\: v_{j_k}=u_{i_k}$}.

Important special cases of join are
crossproduct
$ \relat_1\tapnum{n}\!\CrPr\relat_2\tapnum{m} = \relat_1\tapnum{n}\!\Join_{\emptySet}\relat_2\tapnum{m} $,
intersection
$ \relat_1\tapnum{n}\cap\relat_2\tapnum{n} = \relat_1\tapnum{n}\JOIN{1=1,\dots n=n}\relat_2\tapnum{n} $,
and transducer composition
$ \relat_1\tapnum{2}\circ\relat_2\tapnum{2} = \CProj{2}( \relat_1\tapnum{2}\JOIN{2=1}\relat_2\tapnum{2} )$.

Unfortunately, rational relations are {\em not} closed under arbitrary joins
\cite{kempe+champarnaud+eisner:2004a}.
Since the join operation is very useful in practical applications
(Sec.~\ref{sec:applic}),
it is helpful to have even a partial algorithm:
hence our motivation for studying auto-intersection.

\subsection{Auto-Intersection}\label{sec:op:autoint}

For any distinct $i_1, j_1, \dots$ $i_r, j_r \in \aRange{1,n}$,
we define an {\it auto-intersection} operator
$\AInt{i_1=j_1,i_2=j_2,\dots i_r=j_r}$. It maps a relation
$\relat\tapnum{n}$ to a subset of that relation, preserving tuples
$s\tapnum{n}$ whose elements are equal in pairs as specified, but
removing other tuples from the support of the
relation.\footnote{The requirement that the $2r$ indices be distinct
  mirrors the similar requirement on join and is needed in
  \eqref{eq:AInt-MTInt-Analogy}.  But it can be evaded by
  duplicating tapes: the illegal operation
  $\AInt{1=2,2=3}(\relat)$ can be computed as
  $\CProj{3}(\AInt{1=2,3=4}(\Proj{1,2,2,3}(\relat)))$.}
The formal definition is:
\begin{eqnarray}
\spc{-2ex}
\scalebox{0.86}[1.0]{$
  \left(\AInt{i_1=j_1,\dots i_r=j_r}(\relat\tapnum{n})\right)(\aTuple{s_1,\dots s_n})$}	
  &  \!\DefAs\! &
  \begin{cases}
\scalebox{0.86}[1.0]{$
    \relat\tapnum{n}(\aTuple{s_1,\dots s_n})$} &
\scalebox{0.85}[1.0]{$
	\text{if } (\forall k\!\in\!\aRange{1,r}) s_{i_k}\!=\!s_{j_k} $}\ \ \ 		\\
    \spc{8ex}\srZero & \text{otherwise}
  \end{cases}
	\label{eq:Op:AutoInt}
\end{eqnarray}

It is easy to check that auto-intersecting a relation is different
from joining the relation with its own projections.
Actually, join and auto-intersection are related by the following equalities:
\begin{equation}
\scalebox{0.85}[1.0]{$
  \relat_1\tapnum{n} \JOIN{i_1=j_1,\dots i_r=j_r} \relat_2\tapnum{m}  \; = \;
	\CProj{n+j_1,\dots n+j_r}\left(\;
		\AInt{i_1=n+j_1,\dots i_r=n+j_r}(\;
			\relat_1\tapnum{n} \!\CrPr\! \relat_2\tapnum{m}
		    \;)\;
	\right)
$}
  \label{eq:Op:MltInt:proc}
\end{equation}
\begin{equation}
\spc{-2ex}
\scalebox{0.83}[1.0]{$
  \AInt{i_1=j_1,\dots i_r=j_r}( \relat\tapnum{n} )
	\; = \;
	\relat\tapnum{n}  \JOIN{i_1=1,j_1=2,\dots i_r=2r-1,j_r=2r}
		\left(\underbrace{(\Proj{1,1}(\Sigma^*) \!\CrPr\!
                    \cdots \!\CrPr\! \Proj{1,1}(\Sigma^*)}_{r
                  \text{ times}}\right)
$}
	\label{eq:AInt-MTInt-Analogy}
\end{equation}

Thus, for any class of difficult join instances whose results are
non-rational or have undecidable
properties~\cite{kempe+champarnaud+eisner:2004a},
there is a corresponding class of difficult
auto-intersection instances, and vice-versa.  Conversely, a partial
solution to one problem would yield a partial solution to the other.

An auto-intersection on a single pair of tapes
is said to be a {\it single-pair\/} one. 
An auto-intersection on multiple pairs of tapes
can be defined in terms of multiple single-pair auto-intersections: 
\begin{equation}
  \AInt{i_1=j_1,\dots i_r=j_r}(\; \relat\tapnum{n} \;) \;\DefAs\;
  \AInt{i_r=j_r}(\; \cdots  \AInt{i_1=j_1}(\; \relat\tapnum{n} \;)
  \cdots \;)
	\label{eq:Op:AutoInt:Split}
\end{equation}


\section{Compilation of Auto-Intersection
	\label{sec:alg}}

We now briefly recall a single-pair auto-intersection algorithm
and the class of bounded delay auto-intersections that this algorithm can handle.
For a detailed exposure see \cite{kempe+al:2005a}.

\subsection{Post's Correspondence Problem
	\label{ssec:alg:pcp}}

Unfortunately,
auto-intersection (and hence join) can be reduced to Post's Correspondence Problem (PCP)
\cite{post:1946}.
Actually, any PCP instance can be represented as an unweighted 2-FSM,
and the set of all solutions to the instance equals the auto-intersection of the
2-FSM~\cite{kempe+champarnaud+eisner:2004a}.

Since it can generally not be decided whether any solution exists to an arbitrary PCP instance,
it is also undecidable whether the result of auto-intersection is empty.
Therefore,
no partial auto-intersection algorithm can be ``complete'' in the sense that it
always returns a correct $n$-FSM if it is rational, and
always terminates with an error code otherwise.
Such an algorithm would make PCP generally decidable since
a returned $n$-FSM can always be tested for emptiness, and
an error code indicates non-rationality and hence non-emptiness.

\subsection{A class of rational auto-intersections
	\label{ssec:alg:theta}}

Although there cannot exist a fully general algorithm,
~$A\tapnum{n}\!=\!\AInt{i=j}(A_1\tapnum{n})$
can be compiled for a class of triples $\aTuple{A_1\tapnum{n},i,j}$
whose definition is based on the notion of
{\it delay}~\cite{frougny+sakarovitch:1993,mohri:2003}.
The delay $\delay{i,j}(s\tapnum{n})$
is the difference of length of the strings $s_i$ and $s_j$ of the tuple $s\tapnum{n}$~:
~$\delay{i,j}(s\tapnum{n}) = |s_i|\!-\!|s_j|  \;\;(i,j\!\in\!\aRange{1,n})$.
We call the delay {\it bounded}\/ if its absolute value does not exceed some limit.
The delay of a path $\path\tapnum{n}$ results from its labels on tapes $i$ and $j$:
~$\delay{i,j}(\path\tapnum{n}) = |(\lab(\path\tapnum{n}))_i|\!-\!|(\lab(\path\tapnum{n}))_j|$. \linebreak
A path has bounded delay if all its prefixes have bounded delay,\footnote{
  Any finite path has bounded delay (since its label is of finite length).
  An infinite path (traversing cycles) may have bounded or unbounded delay.
  For example,
  the delay of a path labeled with $(\aTuple{ab,\eps} \aTuple{\eps,xz})^h$ 
  is bounded by 2 for any $h$,
  whereas that of a path labeled with $\aTuple{ab,\eps}^h \aTuple{\eps,xz}^h$ 
  is unbounded for $h\longrightarrow\infty$.
}
and an $n$-WFSM has bounded delay if all its successful paths have bounded delay.

As earlier reported~\cite{kempe+al:2005a},
if an $n$-WFSM $A_1\tapnum{n}$ does not contain a path
traversing both a cycle with positive and a cycle with negative delay
w.r.t. tapes $i$ and $j$,\footnote{
  Note that the $n$-WFSM may have cycles of both types, but not on the same path.
}
then the delay of all paths of its auto-intersection
$A\tapnum{n}\!=\!\AInt{i=j}(A_1\tapnum{n})$
is bounded by some $\delayLimit{i,j}$,
and this bound can be compiled from $A_1\tapnum{n}$.

\subsection{An auto-intersection algorithm
	\label{ssec:alg:algorithm}}

Our algorithm for the above mentioned class of rational auto-intersections
proceeds in three steps~\cite{kempe+al:2005a,kempe+al:2005b}~:

\begin{enumerate}
\item
  Test whether the triple $\aTuple{A_1\tapnum{n},i,j}$ fulfills the above conditions. \\
  \spc{2ex} If not, then the algorithm exits with an error code.

\item
  Calculation of the bound $\delayLimit{i,j}$ for the delay of the auto-intersection \\

  \spc{2ex} $A\tapnum{n}\!=\!\AInt{i=j}(A_1\tapnum{n})$.

  \vspace{0.5ex}

\item
  Construction of the auto-intersection within the bound.
\end{enumerate}

\begin{figure}

  \begin{center}
    \spc{2mm}%
    \begin{minipage}[c]{7mm}
      (a) \\

      $A_1\tapnum{3}$ \\
      \vspace{2mm}
    \end{minipage}%
    \begin{minipage}[c]{35mm}
      \includegraphics[scale=0.40]{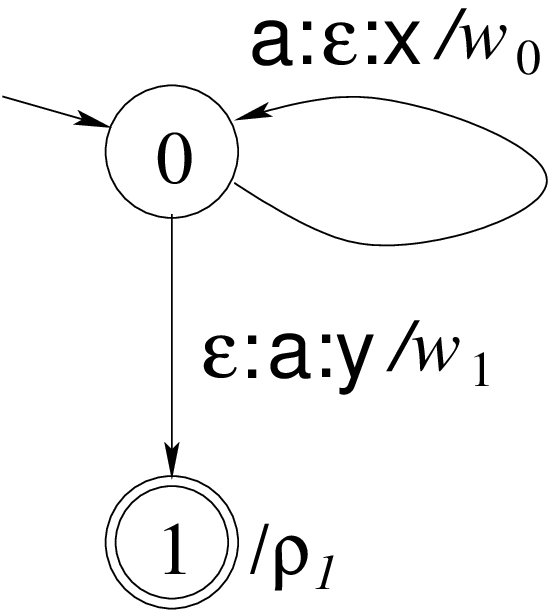}
    \end{minipage}%
    \begin{minipage}[c]{10mm}
      \spc{5mm}(b) \\

      $A\tapnum{3}=\AInt{1=2}\!(A_1\tapnum{3})$ \\
      \vspace{4mm}
    \end{minipage}%
    \begin{minipage}[c]{70mm}
      \includegraphics[scale=0.40]{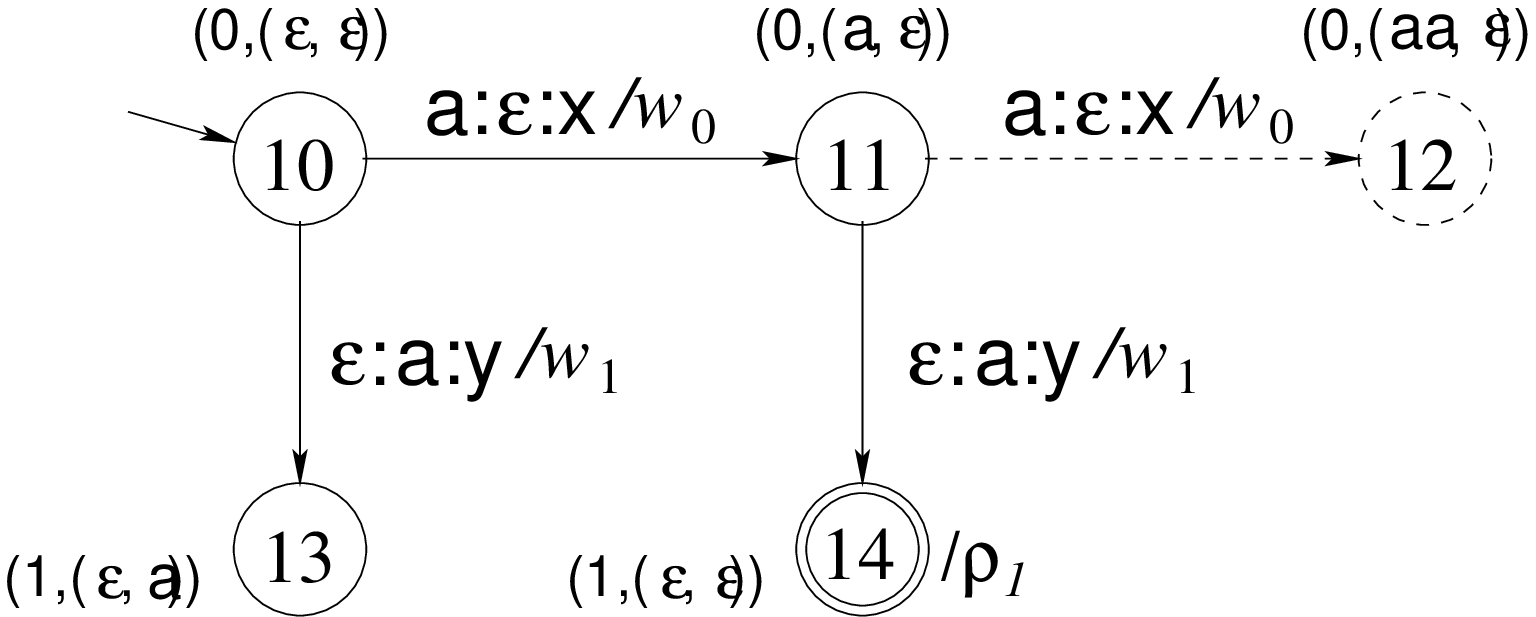}
    \end{minipage}
    \caption{(a) A $3$-WFSM and (b) its auto-intersection
      \label{fig:algorithm}}
  \end{center}

\end{figure}

Figure~\ref{fig:algorithm} illustrates step~3 of the algorithm:~
State~0, the initial state of $A_1\tapnum{3}$, is copied as initial state~10 to $A\tapnum{3}$.
Its annotation, $\aTuple{0,\aTuple{\eps,\eps}}$,
indicates that it is a copy of state~0
and has leftover strings $\aTuple{\eps,\eps}$.
Then, all out-going transitions of state~0 and their target states are copied to $A\tapnum{3}$,
as states~11 and~13.
A transitions is copied with its original label and weight.
The annotation of state~11 indicates that it is a copy of state~0
and has leftover strings $\aTuple{\textsf{a},\eps}$.
These leftover strings result from concatenating the leftover strings of state~10, $\aTuple{\eps,\eps}$,
with the relevant components, $\aTuple{\textsf{a},\eps}$, of the transition label {\sf a:$\eps$:x}.
For each newly created state $q\!\in\!Q_{A}$,
we access the corresponding state $q_1\!\in\!Q_{A_1}$,
and copy $q_1$'s out-going transitions with their target states to $A\tapnum{3}$,
until all states of $A\tapnum{3}$ have been processed.

State~12 is not created because the delay of its leftover strings
$\aTuple{\textsf{aa},\eps}$ exceeds the pre-calculated bound of $\delayLimit{1,2}\!=\!1$.
The longest common prefix of the two leftover strings of a state is removed.
Hence state~14 has leftover strings $\aTuple{\eps,\eps}$
instead of $\aTuple{\textsf{a},\eps}\aTuple{\eps,\textsf{a}}\!=\!\aTuple{\textsf{a},\textsf{a}}$.
A final state is copied with its original weight if it has leftover strings $\aTuple{\eps,\eps}$,
and with weight $\srZero$ otherwise.
Therefore, state~14 is final and state~13 is not.

The construction is proven to be correct and to terminate~\cite{kempe+al:2005a,kempe+al:2005b}.
It can be performed simultaneously on multiple pairs of tapes.


\section{Applications
  \label{sec:applic}}

This section focuses
on demonstrating the augmented descriptive power $n$-WFSMs,
w.r.t. to 1- and 2-WFSMs (acceptors and transducers),
and on exposing the practical importance of the join operation.
It also aims at illustrating how to use $n$-WFSMs,
in practice.
Indeed, some of the applications are not feasible with 1- and 2-WFSMs.
The section does not focus on the presented applications {\em per se}.


\subsection{Morphological Analysis of Semitic Languages
  \label{ssec:semitic}}

$n$-WFSMs have been used in the morphological analysis of Semitic languages
\cite[e.g.]{kay:1987,kiraz:1997,kiraz:2000}.

Table~\ref{tab:semitic} by Kiraz \cite{kiraz:1997}
shows the ``synchronization'' of the quadruple
$ s\tapnum{4} = \aTuple{\textsf{aa}, \textsf{ktb}, \textsf{waCVCVC}, \textsf{wakatab}} $
in a $4$-WFSM representing an Arabic morphological lexicon.
Its first tape encodes a word's vowels,
its second the consonants (representing the root),
its third the affixes and the templatic pattern (defining how to combine consonants and vowels),
and its fourth the word's surface form.

Any of the tapes can be used for input or output.
For example,
for a given root and vowel sequence,
we can obtain all existing surface forms and templates.
For a given root and template,
we can obtain all existing vowel sequences and surface forms, etc.

\begin{table}

  \begin{center}

    \begin{minipage}{60mm}
      \begin{tabular}{*7{c} p{4mm} l} \cline{1-7}
        ~  &    &    & a  &    & a  &    &  & vocalism            \\ \cline{1-7}
        ~  &    & k  &    & t  &    & b  &  & root                \\ \cline{1-7}
        w  & a  & C  & V  & C  & V  & C  &  & pattern and affixes \\ \cline{1-7}
        w  & a  & k  & a  & t  & a  & b  &  & surface form        \\ \cline{1-7}
      \end{tabular}
    \end{minipage}
    \begin{minipage}{52mm}
      \caption{Multi-tape-based morphological anaysis of Arabic;  \newline
        table adapted from Kiraz \cite{kiraz:1997}
        \label{tab:semitic}}
    \end{minipage}

  \end{center}

\end{table}


\subsection{Intermediate Results in Transduction Cascades
  \label{ssec:cascades}}

Transduction cascades have been extensively used in language and speech processing
\cite[e.g.]{ait+chanod:1997,pereira+al:1997,kumar+byrne:2003}.

\begin{figure}[htbp]

  \begin{center}
    \includegraphics[scale=0.40]{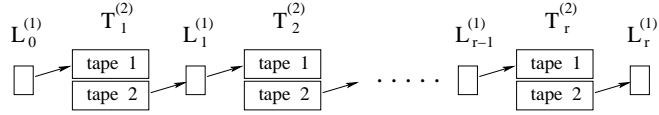}
    
    \caption{Classical $2$-WFSM transduction cascade
      \label{fig:cascade1}}
  \end{center}

\end{figure}

In a classical weighted transduction cascade (Figure~\ref{fig:cascade1}),
consisting of transducers $T_1\tapnum{2} \dots\; T_r\tapnum{2}$,
a weighted input language $L_0\tapnum{1}$,
consisting of one or more words,
is composed with the first transducer, $T_1\tapnum{2}$, on its input tape.
The output projection of this composition is the first intermediate result,
$L_1\tapnum{1}$.
It is further composed with the second transducer, $T_2\tapnum{2}$,
which leads to the second intermediate result, $L_2\tapnum{1}$, etc..
Generally,
$
L_i\tapnum{1} = \Proj{2}( L_{i-1}\tapnum{1} \diamond T_i\tapnum{2} )
\;\;(i\!\in\!\aRange{1,r})
$.
The output projection of the last transducer is the final result, $L_r\tapnum{1}$.

At any point in the cascade, previous intermediate results cannot be accessed.
This holds also if the cascade is composed into a single transducer:
$ T\tapnum{2} = T_1\tapnum{2} \diamond \dots \diamond T_r\tapnum{2} $.
None of the ``incorporated'' sub-relations of $T\tapnum{2}$
can refer to a sub-relation other than its immediate predecessor.

\begin{figure}[htbp]

  \begin{center}
    \includegraphics[scale=0.40]{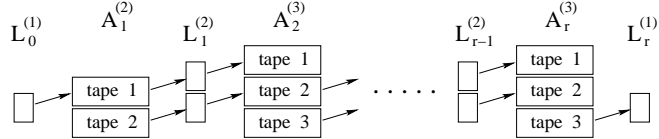}
    
    \caption{$n$-WFSM transduction cascade
      \label{fig:cascade2}}
  \end{center}

\end{figure}

In multi-tape transduction cascade,
consisting of $n$-WFSMs $A_1\tapnum{n_1}\!\dots A_r\tapnum{n_r}$,
any intermediate results can be preserved and used by subsequent transductions.
Figure~\ref{fig:cascade2}
shows an example where two previous results are preserved at each point,
i.e.,
each intermediate result, $L_i\tapnum{2}$, has two tapes.
The projection of the output tape of the last $n$-WFSM is the final result,
$L_r\tapnum{1}$~:
\begin{eqnarray}
  L_1\tapnum{2}  & = &
	L_0\tapnum{1} \JOIN{1=1} A_1\tapnum{2}	\\
  L_i\tapnum{2}  & = &
	\Proj{2,3}(\; L_{i-1}\tapnum{2} \JOIN{1=1,2=2} A_i\tapnum{3} \;)
	\spc{10ex} (i\!\in\!\aRange{2, r-1})	\\
  L_r\tapnum{1}  & = &
	\Proj{3}(\; L_{r-1}\tapnum{2} \JOIN{1=1,2=2} A_r\tapnum{3} \;)
\end{eqnarray}

This augmented descriptive power is also available
if the whole cascade is joined into a single $2$-WFSM, $A\tapnum{2}$,
although $A\tapnum{2}$ has only two tapes (in this example),
for input and output, respectively.
$A\tapnum{2}$ can be iteratively constructed
(Any $B_i\tapnum{m}$ is the join of $A_1\tapnum{2}$ to $A_i\tapnum{3}$)~:
\begin{eqnarray}
  B_1\tapnum{2} & = &
	A_1\tapnum{2}   \\
  B_i\tapnum{3} & = &
	\Proj{1,n-1,n}(\; B_{i-1}\tapnum{m} \JOIN{n-1=1,n=2} A_i\tapnum{3} \;)
	\spc{4ex} (i\!\in\!\aRange{2,r} \,,\; m\!\in\!\aSet{2,3})  \\
  A\tapnum{2}	& = &
	\Proj{1,n}(\; B_r \;)
\end{eqnarray}

\noindent
Each (except the first) of the ``incorporated'' multi-tape sub-relations in $A\tapnum{2}$
will still refer to its two predecessors.

\subsection{Induction of Morphological Rules
  \label{ssec:morph}}

Induction of morphemes and morphological rules from corpora,
both supervised and unsupervised,
is a subfield of NLP on its own
\cite[e.g.]{brent:1999,goldsmith:2001,creutz+al:2007}.
We do not propose a new method for inducing rules,
but rather demonstrate how known steps can be conveniently performed
in the framework of $n$-ary relations.

Learning morphological rules from a raw corpus can include, among others:
(1) generating the least costly rule for a given word pair, that rewrites one word to the other,
(2) identifying the set of pairs over all corpus words where a given rule applies, and
(3) rewriting a given word by means of one or several rules.

\subsubsection{Construction of a rule generator
  \label{sssec:rule-gen}}

For any word pair, such as $\aTuple{parl{\bf er}, parl{\bf ons}}$
(French, {\em [to] speak}, {\em [we] speak}),
the generator shall provide a rule, such as ``{\sf .er:ons}'',
suitable for rewriting the first to the second word at minimal cost.
In a rule, a dot shall mean that one or more letters remain unmodified,
and an {\sf x:y}-part that substring {\sf x} is replaced by substring {\sf y}.

\noindent
We begin with a $4$-WFSM that defines rewrite operations:
\begin{equation}
  \scalebox{0.82}[1]{$ %
    A_1\tapnum{4} = \left(
    \aTuple{\aTuple{\Any,\Any, {\rm{\LARGE\bf .}}\,,\ocK}_{\aSet{1=2}},0}
    \cup 
    \aTuple{\aTuple{\Any,\eps,\Any,\ocD}_{\aSet{1=3}},0}
    \cup 
    \aTuple{\aTuple{\eps,\Any,\Any,\ocI}_{\aSet{2=3}},0}
    \cup 
    \aTuple{\aTuple{\eps,\eps, {\rm{\Large\bf :}}\,,\ocS},0}
    \right)^* $ %
    \label{eq-morph-a1}
  }
\end{equation}

\noindent
where $\Any$ ~can be instantiated by any symbol,
$\eps$ is the empty string,
$_{\aSet{i=j}}$ a constraint
requiring the $\Any$'s on tapes $i$ and $j$ to be instantiated by the same symbol
\cite{nicart+al:2006a},\footnote{
  Deviating from ~\cite{nicart+al:2006a},
  we denote symbol constraints
  similarly to join and auto-intersection constraints.
}
and $0$ a weight over the tropical semiring.

\begin{figure}

  \begin{center}

    \begin{minipage}[b]{45mm}
      \begin{center}
        \includegraphics[scale=0.35]{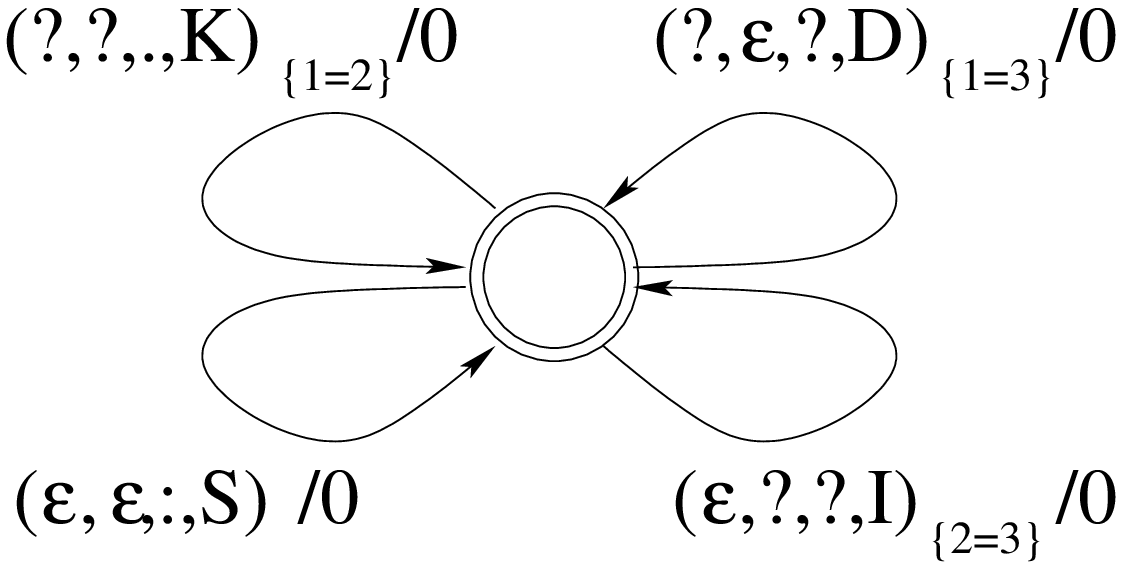}
        
        \begin{minipage}{38mm}
          \caption{Initial form $A_1\tapnum{4}$ of the rule generator
            \label{fig-morph-a1}}
        \end{minipage}
      \end{center}
    \end{minipage}
    \spc{5mm}
    \begin{minipage}[b]{65mm}
      \begin{center}
        {\footnotesize
          \begin{tabular}{c p{3mm} |*7{c}|} \cline{3-9}
            {\it word 1}                &  & s & w & u & m &   &   &    \\ \cline{3-9}
            {\it word 2}                &  & s & w &   &   &   & i & m  \\ \cline{3-9}
            {\it preliminary rule}      &  & . & . & u & m & : & i & m  \\ \cline{3-9}
            {\it preliminary op. codes} &  & \ocK & \ocK & \ocD & \ocD & \ocS & \ocI & \ocI  \\ \hline\hline
            {\it final rule}            &  & . &   & u & m & : & i & m  \\ \cline{3-9}
            {\it final operation codes} &  & \ocK & \ock & \ocD & \ocd & \ocS & \ocI & \oci  \\ \cline{3-9}\cline{3-9}
            {\it weights}               &  & 1 & 0 & 4 & 2 & 0 & 4 & 2  \\ \cline{3-9}\cline{3-9}
          \end{tabular}%
        }
        
        \vspace{1mm}
        \begin{minipage}{52mm}
          \caption{Mapping from the word pair $\aTuple{swum, swim}$
            to various sequences
            \label{tab-morph-a1}}
        \end{minipage}
      \end{center}
    \end{minipage}
    
  \end{center}

\end{figure}

Figure~\ref{fig-morph-a1} shows the graph of $A_1\tapnum{4}$
and Figure~\ref{tab-morph-a1} (rows~1--4) the purpose of its tapes:
Tapes 1 and 2 accept any word pair,
tape 3 generates a preliminary form of the rule,
and tape 4 generates a sequence of preliminary operation codes.
The following four cases can occur when $A_1\tapnum{4}$ reads a word pair
(cf. Eq.~\ref{eq-morph-a1})~:
\begin{enumerate}
\item
  $\aTuple{\Any,\Any,{\LARGE\rm{\bf .}}\,,\ocK}_{\aSet{1=2}}$:
  two identical letters are accepted, meaning a letter is kept from word 1 to word 2,
  which is represented by a ``\rm{\bf .}'' in the rule
  and \ocK ~(keep) in the operation codes,
\item
  $\aTuple{\Any,\eps,\Any,\ocD}_{\aSet{1=3}}$:
  a letter is deleted from word 1 to 2,
  expressed by this letter in the rule
  and \ocD ~(delete) in the operation codes,
\item
  $\aTuple{\eps,\Any,\Any,\ocI}_{\aSet{2=3}}$:
  a letter is inserted from word 1 to 2,
  expressed by this letter in the rule
  and \ocI ~(insert) in the operation codes
\item
  $\aTuple{\eps,\eps,{\large\rm{\bf :}}\,,\ocS}$:
  no letter is matched in either word,
  a ``{\bf :}'' is inserted in the rule,
  and a \ocS ~(separator) in the operation codes.
\end{enumerate}

Next, we compile $C_1\tapnum{1}$ that constrains the order of operation codes.
For example,
~\ocD ~must be followed by \ocS,
~\ocI ~must be preceded by \ocS, etc.
The constraints are enforced through join
(Fig.~\ref{tab-morph-a1} row~4)~:
$A_2\tapnum{4} = A_1\tapnum{4} \JOIN{4=1} C\tapnum{1}$.

Then, we create $B_1\tapnum{2}$ that maps temporary rules to their final form
by replacing a sequence of dots (longest match) by a single dot.
We join $B_1\tapnum{2}$ with the previous result
(Fig.~\ref{tab-morph-a1} rows~3,~5)~:
$A_3\tapnum{5} = A_2\tapnum{4} \JOIN{3=1} B_1\tapnum{2}$.

Next, we compile $B_2\tapnum{2}$
that creates more fine-grained operation codes.
In a sequence of equal capital letters, it replaces each but the first one
with its small form.
For example, {\sf DDD} becomes {\sf Ddd}.
$B_1\tapnum{2}$ is joined with the previous result
(Fig.~\ref{tab-morph-a1} rows~4,~6)~:
$A_4\tapnum{6} = A_3\tapnum{5}  \JOIN{4=1} B_2\tapnum{2}$.

$C_1\tapnum{1}$, $B_1\tapnum{2}$, and $B_2\tapnum{2}$
can be compiled as unweighted automata with a tool such as {\sc Xfst}
\cite{karttunen+al:1998,beesley+karttunen:2003}
and then be enhanced with neutral weights.

Finally, we assigns weights to the fine-grained operation codes
by joining
$
B_3\tapnum{1} = (
\aTuple{\ocK,1} \cup \aTuple{\ock,0} \cup
\aTuple{\ocD,4} \cup \aTuple{\ocd,2} \cup
\aTuple{\ocI,4} \cup \aTuple{\oci,2} \cup
\aTuple{\ocS,0} )^*
$
with the previous result
(Fig.~\ref{tab-morph-a1} rows 6,~7)~:
$ A_5\tapnum{6} = A_4\tapnum{6} \JOIN{6=1} B_3\tapnum{1} $.

We keep only the tapes of the word pair and of the final rule
in the generator
(Fig.~\ref{tab-morph-a1} rows~1,~2,~5).
All other tapes are of no further use:
\begin{equation}
  G\tapnum{3} = \Proj{1,2,5} \left( A_5\tapnum{6} \right)
\end{equation}

The rule generator $G\tapnum{3}$
maps any word pair to a finite number of rewrite rules
with different weight, expressing the cost of edit operations.
The optimal rule (with minimal weight)
can be found through $n$-tape best-path search
\cite{kempe:2009b}.

\subsubsection{Using rewrite rules
  \label{sssec:rule-use}}

We suppose that the rules generated from random word pairs
undergo some statistical selection process that aims at retaining only meaningful rules.

To facilitate the following operations,
a rule's representation can be  \linebreak changed
from a string, such as $s\tapnum{1}\!=$``{\sf .er:ons}'',
to a 2-WFSM $r\tapnum{2}$ encoding the  \linebreak same relation.
This is done by joining the rule with the generator:  \linebreak
$ r\tapnum{2} = \Proj{1,2}\left( G\tapnum{3}\JOIN{3=1}s\tapnum{1} \right) $.
An $r\tapnum{2}$ resulting from ``{\sf .er:ons}'',
accepts (on tape 1) only words ending in ``{\sf er}''
and changes (on tape 2) their suffix to ``{\sf ons}''.

Similarly, a 2-WFSM $R\tapnum{2}$ that encodes all selected rules
can be generated by joining the set of all rules (represented as strings) $S\tapnum{1}$
with the generator:
$ R\tapnum{2} = \Proj{1,2}\left(\; G\tapnum{3} \JOIN{3=1} S\tapnum{1} \;\right) $.

To find all pairs $P\tapnum{2}$ of words from a corpus where a particular rule applies,
we compile the automaton $W\tapnum{1}$ of all corpus words,
and compose it on both tapes of $r\tapnum{2}$~:
$ P\tapnum{2} = W\tapnum{1} \circ r\tapnum{2} \circ  W\tapnum{1} $.~
Similarly, identifying all word pairs ${P^\prime}\tapnum{2}$
over the whole corpus where any of the rules applies
(i.e., the set of ``valid'' pairs)
can be obtained through:
$ {P^\prime}\tapnum{2} = W\tapnum{1} \circ R\tapnum{2} \circ  W\tapnum{1} $

Rewriting a word $w\tapnum{1}$ with a single rule $r\tapnum{2}$ is done by
$ w_2\tapnum{1} = \Proj{2} ( w_1\tapnum{1} \circ\, r\tapnum{2} ) $
and
$ w_1\tapnum{1} = \Proj{1} ( r\tapnum{2} \circ  w_2\tapnum{1} ) $.~
Similarly, rewriting a word $w\tapnum{1}$ with all selected rules is done by
$ W_2\tapnum{1} = \Proj{2} ( w_1\tapnum{1} \circ R\tapnum{2} ) $
and
$ W_1\tapnum{1} = \Proj{1} ( R\tapnum{2} \circ  w_2\tapnum{1} ) $.


\subsection{String Alignment for Lexicon Construction
  \label{ssec:lexalign}}

Suppose, we want to create a (non-weighted) transducer, $D\tapnum{2}$,
from a list of word pairs $s\tapnum{2}$
of the form $\aTuple{\WordD{inflected form}, \WordD{lemma}}$,
e.g., $\aTuple{\Word{swum}, \Word{swim}}$,
such that each path of the transducer is labeled with one of the pairs.
We want to use only transition labels of the form
$\aTuple{\sigma,\sigma}$, $\aTuple{\sigma,\eps}$,
or $\aTuple{\eps,\sigma}$ ~($\forall\sigma\in\Sigma$),
while keeping paths as short as possible.
For example,
$\aTuple{\Word{swum}, \Word{swim}}$ should be encoded either by the sequence
$\aTuple{\Word{s},\Word{s}}\aTuple{\Word{w},\Word{w}}%
\aTuple{\Word{u},\eps}\aTuple{\eps,\Word{i}}\aTuple{\Word{m},\Word{m}}$
or by
$\aTuple{\Word{s},\Word{s}}\aTuple{\Word{w},\Word{w}}%
\aTuple{\eps,\Word{i}}\aTuple{\Word{u},\eps}\aTuple{\Word{m},\Word{m}}$,
rather than by the ill-formed
$\aTuple{\Word{s},\Word{s}}\aTuple{\Word{w},\Word{w}}%
\aTuple{\Word{u},\Word{i}}\aTuple{\Word{m},\Word{m}}$,
or the sub-optimal
$\aTuple{\Word{s},\eps}\aTuple{\Word{w},\eps}\aTuple{\Word{u},\eps}\aTuple{\Word{m},\eps}$
$\aTuple{\eps,\Word{s}}\aTuple{\eps,\Word{w}}\aTuple{\eps,\Word{i}}\aTuple{\eps,\Word{m}}$.

We start with a $5$-WFSM over the real tropical semiring
\cite{isabelle+kempe:2004}~:
\begin{equation}
\spc{-2ex}\scalebox{0.82}[0.95]{
  $
  A_1\tapnum{5} \;=\; \left(\;
  \aTuple{\aTuple{\Any,\Any,\Any,\Any,\ocK}_{\aSet{1=2=3=4}} , 0}
  \,\cup\, \aTuple{\aTuple{\eps,\Any,\algnEps,\Any,\ocI}_{\aSet{2=4}} , 1}
  \,\cup\, \aTuple{\aTuple{\Any,\eps,\Any,\algnEps,\ocD}_{\aSet{1=3}} , 1} \;\right)^*
  \label{eq:align}
  $
}
\end{equation}

\noindent
where $\algnEps$ is a special symbol representing $\eps$ in an alignment,
$_{\aSet{1=2=3=4}}$ a constraint
requiring the $\Any$'s on tapes $1$ to $4$ to be instantiated
by the same symbol~\cite{nicart+al:2006a},
and $0$ and $1$ are weights.

\begin{figure}

  \begin{center}
    
    \begin{minipage}[b]{50mm}
      \begin{center}
        \includegraphics[scale=0.38]{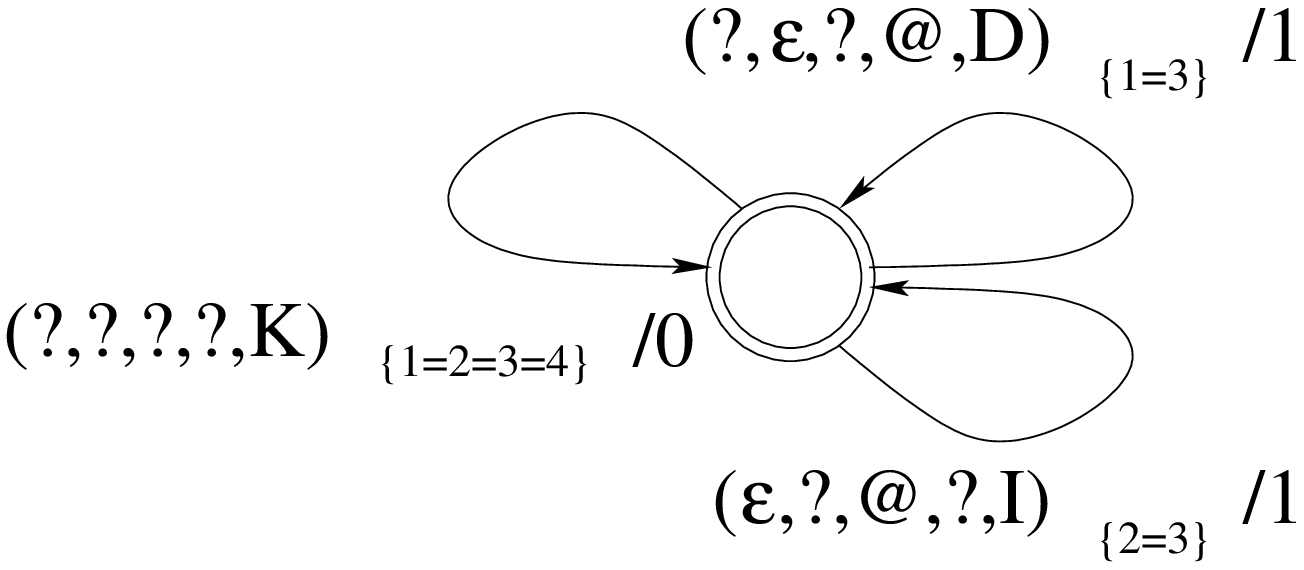}
        
        \begin{minipage}{40mm}
          \caption{Initial form $A_1\tapnum{5}$ of a word pair aligner
            \label{fig-lexalign-a1}}
        \end{minipage}
      \end{center}
    \end{minipage}
    \spc{10mm}
    \begin{minipage}[b]{45mm}
      \begin{center}
        \begin{tabular}{c p{5mm} |*5{c}|} \cline{3-7}
          {\it input word 1}     &  & s & w & u &   & m  \\ \cline{3-7}
          {\it input word 2}     &  & s & w &   & i & m  \\ \cline{3-7}
          {\it output word 1}    &  & s & w & u & @ & m  \\ \cline{3-7}
          {\it output word 2}    &  & s & w & @ & i & m  \\ \cline{3-7}
          {\it operation codes}  &  & \ocK & \ocK & \ocD & \ocI & \ocK  \\ \cline{3-7}
          {\it weights}          &  & 0 & 0 & 1 & 1 & 0  \\ \cline{3-7}
        \end{tabular}
        \begin{minipage}{37mm}
          \caption{Alignment of the word pair $\aTuple{swum, swim}$
            \label{tab-lexalign-a1}}
        \end{minipage}
      \end{center}
    \end{minipage}
    
  \end{center}

\end{figure}

Figure~\ref{fig-lexalign-a1} shows the graph of $A_1\tapnum{5}$
and Figure~\ref{tab-lexalign-a1} (rows~1--5) the purpose of its tapes:
Input word pairs $s\tapnum{2}\!=\!\aTuple{s_1,s_2}$ will be matched on tape 1 and 2,
and aligned output word pairs generated from tape 3 and 4.
A symbol pair $\aTuple{\Any,\Any}$ read on tape 1 and 2
is identically mapped to $\aTuple{\Any,\Any}$ on tape 3 and 4,
a $\aTuple{\eps,\Any}$ is mapped to $\aTuple{\algnEps,\Any}$,
and a $\aTuple{\Any,\eps}$ to $\aTuple{\Any,\algnEps}$.
$A_1\tapnum{5}$ will introduce $\algnEps$'s in $s_1$ (resp. in $s_2$) at positions
where $D\tapnum{2}$ shall have $\aTuple{\eps,\sigma}$-
(resp. a $\aTuple{\sigma,\eps}$-) transitions.\footnote{%
  Later, we simply replace in $D\tapnum{2}$ all $\algnEps$ by $\eps$.
}
Tape~5 generates a sequence of operation codes:
\ocK ~(keep), \ocD ~(delete), \ocI ~(insert).
For example, $A_1\tapnum{5}$ will map $\aTuple{\Word{swum},\Word{swim}}$,
among others,
to $\aTuple{\Word{swu\algnEps m}, \Word{sw\algnEps im}}$ with \ocK\ocK\ocD\ocI\ocK~
and to $\aTuple{\Word{sw\algnEps um}, \Word{swi\algnEps m}}$ with \ocK\ocK\ocI\ocD\ocK.

To remove redundant (duplicated) alignments,
we prohibit an insertion to be immediately followed by a deletion,
via the constraint:
$
C\tapnum{1} =
( \ocK \cup \ocI \cup \ocD )^* - ( \anySym^* \;\ocI \;\ocD \;\anySym^* )
$.
The constraint is imposed through join and the operations tape is removed:
\begin{equation}
  \textit{Aligner}\tapnum{4} = \CProj{5}\!\left(\; A_1\tapnum{5} \JOIN{5=1} C\tapnum{1} \;\right)
\end{equation}

The $\textit{Aligner}\tapnum{4}$ will map $\aTuple{\Word{swum},\Word{swim}}$
among other still to $\aTuple{\Word{swu\algnEps m}, \Word{sw\algnEps im}}$
but no to $\aTuple{\Word{sw\algnEps um}, \Word{swi\algnEps m}}$.
The best alignment (with minimal weight)
can be found through $n$-tape best-path search
\cite{kempe:2009b}.


\subsection{Acronym and Meaning Extraction
  \label{ssec:acronyms}}

\newcommand{\mtchltr}[1]{\underline{#1}}

\newcommand{\mtchtxt}[1]{{\small\sf #1}}

The automatic extraction of acronyms and their meaning from corpora
is an important sub-task of text mining,
and received much attention
\cite[e.g.]{yeates+al:2000,pustejovsky+al:2001,schwartz+hearst:2003}.

It can be seen as a special case of string alignment
between a text chunk and an acronym.
For example, the chunk ``\mtchtxt{they have many hidden Markov models}''
can be aligned with the acronym ``\mtchtxt{HMMs}'' in different ways,
such as
``\mtchtxt{they have many \mtchltr{h}idden \mtchltr{M}arkov \mtchltr{m}odel\mtchltr{s}}''
or
``\mtchtxt{t\mtchltr{h}ey have \mtchltr{m}any hidden \mtchltr{M}arkov model\mtchltr{s}}''.
Alternative alignments have different cost,
and ideally the least costly one should give the correct meaning.

An alignment-based approach
can be implemented by means of a 3-WFSM
that reads a text chunk on tape 1 and an acronym on tape 2,
and generates all possible alignments on tape 3,
inserting dots to mark letters used in the acronym.
For the above example this would give ``\mtchtxt{they have many .hidden .Markov .model.s}'',
among others.

The $3$-WFSM can be generated from $n$-ary regular expressions
that define the task in as much detail as required
(cf. Sec.~\ref{ssec:morph} and~\ref{ssec:lexalign}).
For a detailed description see \cite{kempe:2006b}.
The best alignment,
i.e., the most likely meaning of an acronym
is found through $n$-tape best-path search
\cite{kempe:2009b}.

The advantage of aligning via a $n$-WFSM rather than a classical alignment matrix
\cite{wagner+fischer:1974,pirkola+al:2003}
is that the $n$-WFSM can be built from regular expressions that define very subtle criteria,
such as disallowing certain alignments or favoring others
based on weights that depend on long-distance context.


\subsection{Cognate Search
  \label{ssec:cognates}}

Extracting cognates with equal meaning from an English-German dictionary $\text{EG}\tapnum{3}$
that encodes triples 
$ \aTuple{\textit{English word}, \textit{German word}, \textit{part of speech}} $,
means to identify all paths of $\text{EG}\tapnum{3}$ that have similar strings on tapes 1 and 2.

We create a similarity automaton $S\tapnum{2}$ that describes through weights
the degree of similarity between English and German words.
This can either be expressed through edit distance
(cf. Sec.~\ref{ssec:morph}, \ref{ssec:lexalign}, and~\ref{ssec:acronyms})
or through weighted synchronic grapheme correspondences
(e.g.: {\em d}-{\em t}, {\em ght}-{\em cht}, {\em th}-{\em d}, {\em th}-{\em ss}, $\ldots$)~:
$
S\tapnum{2} = \left(
\aTuple{\aTuple{\anySym,\anySym}_{\aSet{1=2}},w_0} \cup
\aTuple{\aTuple{{\it d},{\it t}},w_1} \cup
\aTuple{\aTuple{{\it ght},{\it cht}},w_2} \cup \dots
\right)^*
$

\noindent
When recognizing an English-German word pair,
$S\tapnum{2}$ accepts either any two equal symbols in the two words
(via $\aTuple{\anySym,\anySym}_{\aSet{1=2}}$)
or some English sequence and its German correspondence (e.g. {\em ght} and {\em cht})
with some weight.

The set of cognates $\text{EG}_{\text{cog}}\tapnum{3}$ is obtained by
joining the dictionary with the similarity automaton:
$ \text{EG}_{\text{cog}}\tapnum{3} = \text{EG}\tapnum{3} \JOIN{1=1,2=2} S\tapnum{2} $

\noindent
$\text{EG}_{\text{cog}}\tapnum{3}$
contains all (and only) the cognates with equal meaning in $\text{EG}\tapnum{3}$
such as
$\aTuple{\text{daughter}, \text{tochter}, \text{noun}}$,
$\aTuple{\text{eight}, \text{acht}, \text{num}}$,
or $\aTuple{\text{light}, \text{leicht}, \text{adj}}$.
Weighs of triples express similarity of words.

Note that this result cannot be achieved through ordinary transducer composition.
For example, composing $S\tapnum{2}$ with the English and the German words separately:
$ \Proj{1}(\text{EG}\tapnum{3}) \diamond S\tapnum{2} \diamond \Proj{2}(\text{EG}\tapnum{3}) $,
also yields false cognates such as
$\aTuple{\text{become}, \text{bekommen}}$ ({\em [to] obtain}).


\section{Conclusion
  \label{sec:conclusion}}

The paper recalled
basic definitions about $n$-ary weighted relations and their $n$-WFSMs,
central operations on these relations and machines,
and an algorithm for the important auto-intersection operation.

It investigated the potential of $n$-WFSMs,
w.r.t. classical 1- and 2-WFSMs (acceptors and transducers),
in practical tasks.
Through a series of applications,
it exposed their augmented descriptive power
and the importance of the join operation.
Some of the applications are not feasible with 1- or 2-WFSMs.

In the morphological analysis of Semitic languages, $n$-WFSMs have been used 
to synchronize the vowels, consonants, and
templatic pattern into a surface form.
In transduction cascades consisting of $n$-WFSMs,
 intermediate result can be preserved and used by subsequent transductions.
$n$-WFSMs permit not only to map strings to strings
or string $m$-tuples to $k$-tuples,
but $m$-ary to $k$-ary string relations,
such as an non-aligned word pair to its aligned form,
or to a rewrite rule suitable for mapping one word to the other.
In string alignment tasks,
an $n$-WFSM provides better control over the alignment process than a classical alignment matrix,
since it can be compiled from regular expressions defining very subtle criteria,
such as long-distance dependencies for weights.

\bibliographystyle{plain}

\bibliography{kempe09a}

\end{document}